\documentclass{appolb}
\usepackage{epsfig}

\newcommand{\nc}{\newcommand}
\nc{\postscript}[2] 
{\setlength{\epsfxsize}{#2\hsize}\centerline{\epsfbox{#1}}}
\nc{\bg}{B. Grzadkowski}
\nc{\non}{\nonumber}
\nc{\barx}{\bar{x}}\nc{\pbarn}{\;\hbox {pb}}\nc{\fbarn}{\;\hbox {fb}}
\nc{\veff}{V_{\rm eff}}
\nc{\vtrue}{v_0}
\nc{\vtree}{v}
\nc{\hc}{\hbox {h.c.}} 
\nc{\re}{\hbox {Re}} 
\nc{\im}{\hbox {Im}}
\nc{\mev}{\hbox {MeV}} 
\nc{\gev}{\;\hbox {GeV}} 
\nc{\tev}{\;\hbox {TeV}}
\def\gesim{\lower0.5ex\hbox{$\:\buildrel >\over\sim\:$}} 
\def\lesim{\lower0.5ex\hbox{$\:\buildrel <\over\sim\:$}} 
\nc{\xprd}[3]{{\it Phys.\ Rev.}\ {{\bf D{#1}} (#2), #3}}
\nc{\xprb}[3]{{\it Phys.\ Rev.}\ {{\bf B{#1}} (#2), #3}}
\nc{\xprl}[3]{{\it Phys.\ Rev.\ Lett.}\ {{\bf {#1}} (#2), #3}}
\nc{\pr}[3]{{\it Phys.\ Rep.}\ {{\bf {#1}} (#2), #3}}
\nc{\plb}[3]{{\it Phys.\ Lett.}\ {{\bf B{#1}} (#2), #3}}
\nc{\npb}[3]{{\it Nucl.\ Phys.}\ {{\bf B{#1}} (#2), #3}}
\nc{\ptp}[3]{{\it Prog.\ Theor.\ Phys.}\ {{\bf {#1}} (#2), #3}}
\nc{\zfp}[3]{{\it Z.\ Phys.}\ {{\bf C{#1}} (#2), #3}}
\nc{\mpla}[3]{{\it Mod.\ Phys.\ Lett.}\ {{\bf A{#1}} (#2), #3}}
\nc{\xrmp}[3]{{\it Rev.\ Mod.\ Phys.}\ {{\bf {#1}} (#2), #3}}
\nc{\ijmpa}[3]{{\it Int.\ J.\ Mod.\ Phys.}\
               {{\bf A{#1}} (#2), #3}}
\nc{\jhep}[3]{{\it JHEP}\ {{\bf #1} (#2), #3}}

\nc{\lspace}{\;\;\;\;\;\;\;\;\;\;}  \nc{\llspace}{\lspace \lspace}

\nc{\beq}{\begin{equation}}   \nc{\eeq}{\end{equation}}
\nc{\bea}{\begin{eqnarray}}   \nc{\eea}{\end{eqnarray}}
\nc{\baa}{\begin{array}}      \nc{\eaa}{\end{array}}
\nc{\bit}{\begin{itemize}}    \nc{\eit}{\end{itemize}}
\nc{\ben}{\begin{enumerate}}  \nc{\een}{\end{enumerate}}
\nc{\bce}{\begin{center}}     \nc{\ece}{\end{center}}

\nc{\mh}{m_h}
\nc{\mt}{m_t}
\nc{\mz}{m_Z}
\nc{\la}{\lambda}
\nc{\La}{\Lambda}

\def\half{\frac12}
\def\lcal{{\cal L}}
\def\up#1{^{(#1)}}
\def\inv#1{\frac1{#1}}
\def\ocal{{\cal O}}
\def\pb{\bar\varphi}

   \def\thebibliography#1{\centerline{REFERENCES}
     \list{[\arabic{enumi}]}{\settowidth\labelwidth{[#1]}\leftmargin
     \labelwidth\advance\leftmargin\labelsep\usecounter{enumi}}
     \def\newblock{\hskip .11em plus .33em minus -.07em}\sloppy
     \clubpenalty4000\widowpenalty4000\sfcode`\.=1000\relax}
   
\begin{document}
\pagestyle{plain}
\eqsec
\newcount\eLiNe\eLiNe=\inputlineno\advance\eLiNe by -1
\title{
\vspace{-5cm}
\begin{flushright}
\parbox{4cm}{IFT-31/2001\\
                         UCRHEP-T321\\
March, 2001}
\end{flushright}
\vspace{+4cm}
%
Triviality and Vacuum Stability Bounds on the Higgs Boson Mass beyond the Standard Model%
\thanks{Presented at The XXV International School  of Theoretical Physics,  Ustro\'n, Poland, September 
10-16.2001}%
}
\author{
Bohdan Grzadkowski\footnote{E-mail address: \tt bohdan.grzadkowski@fuw.edu.pl}
\address{Institute of Theoretical Physics, Warsaw University,\\
 Ho\.za 69, PL-00-681 Warsaw, POLAND}
\and
Jos\'e Wudka\footnote{E-mail address: \tt jose.wudka@ucr.edu}
\address{Department of Physics, University of California-Riverside \\
         California 92521-0419, USA}}
\maketitle

\begin{abstract}
The triviality and vacuum stability bounds on the Higgs-boson mass were revisited
in presence of  weakly-coupled new interactions  parameterized
in a model-independent way by effective operators of dimension 6.
It was shown that for the scale of new physics in the region $\La \simeq .5 \div 50 \tev$
the Standard Model triviality upper bound remains unmodified
whereas it is natural to expect that 
the lower bound derived from the requirement of vacuum stability is
 increased
by $40\div60\gev$ depending on the scale $\La$ and strength
of coefficients of effective operators. It turns out that if the Higgs-boson mass
is close to its lower LEP limit then the scale of new physics that
follows from the vacuum stability requirement would be
decreased dramatically even for modest values of coefficients of effective 
operators implying new physics already at the scale of a few $\tev$.
\end{abstract}
\PACS{14.80.Bn, 14.80.Cp}

\newpage
\section{Introduction}
\label{sect:intro}

In spite of a huge experimental effort, the Higgs particle, the last missing ingredient of 
the Standard Model (SM) of electroweak interactions has not been discovered yet. 
For a Higgs-boson mass $\mh \lesim 115 \gev$ the most promising production channel
has been the radiation off a $Z$-boson at LEP2: $e^+e^- \to Z h$; using this reaction
the  LEP data provided a limit~\cite{higgs_limit} on the SM Higgs-boson mass:
$ \mh > 113.2 \gev$.
The Higgs particle also contributes radiatively to several well measured
quantities, this can be used to derive an
upper bound~\cite{prec_data} on $m_h$: $\mh \lesim 212 \gev$ at 95 \% C.L.. 
However, one should be aware that both limits are highly model-dependent.

There exist other theoretical restrictions of $\mh$ based on the so-called
triviality and vacuum stability arguments. As it is well know~\cite{triviality} 
the renormalized $\phi^4$ theory cannot contain an interaction term ($\la \phi^4$) for
any non-zero scalar mass: the theory must be trivial. Within a 
perturbative approach the statement corresponds to the fact
that for any non-zero scalar mass
(since the mass is $ \propto \sqrt{\la}$ this condition corresponds to a
non-vanishing  initial value for the renormalization group (RG)
evolution of $\la$) there exists a finite energy scale at which $\la$ diverges 
(the Landau pole). Consequently, only  for zero scalar mass the
theory can be consistent for all energy scales. An analogous effect 
occurs in the scalar sector of the SM (modified to some extend by the presence
of gauge and Yukawa interactions). This, however,
does not necessarily implies zero Higgs-boson mass since there is no reason
to believe the SM is valid at arbitrarily high energy scale. For
example, it is often assumed that the SM represents the
low energy limit of some underlying more fundamental theory whose heavy
excitations decouple at low energy, but become manifest at a scale $\La$.
Within that scenario the SM is an effective theory valid possibly only at the
energy scale of the order of the Fermi scale: $G_F^{-1/2}\simeq 300\gev $. 

If the SM is to be accurate for energies below $ \La $ the Landau pole
should occur at scale $ \La $ or above, and this condition gives a
($\La$-dependent) upper bound on $ \mh $~\cite{triv_bounds}. 
On the other hand, for sufficiently small $ \mh $ radiative corrections can
destabilize the ground state. This occurs if the scalar
self coupling constant $ \lambda $ becomes negative  at some scale that
can be identified with the scale of new physics $\La$. Alternatively
requiring the SM vacuum to be stable for scales below $ \La$ implies a
lower bound on $ \mh $~\cite{vacuum_bounds}. 

The consequences of the above arguments (triviality and vacuum stability) are 
usually discussed assuming SM interactions. However,
if the scale of new physics is sufficiently low (of the order of a few TeV)
one could expect that the non-standard interactions would modify 
the electroweak theory at the lower scale and influence the scalar potential 
in such a way that the above bounds on the Higgs-boson mass are changed.
The problem deserves a special attention in the context of possible Higgs-boson
discovery~\cite{higgs_disc} at LEP2 at the mass $\mh\simeq 115 \gev$ since
in this case
the SM constraint from vacuum stability
requires $ \Lambda \lesim {\cal O}(100) \tev $~\cite{quiros} (the precise number depends on the
top quark mass) with the attractive possibility
that $ \La$ is actually much lower.

It then becomes interesting to determine the manner in which heavy
physics with scales in the $ 10 \tev $ region modify the stability and
triviality bounds on the Higgs-boson mass. In this lecture we address this 
question in a model-independent way. We parameterize the heavy physics
effects using  an effective Lagrangian satisfying the SM gauge
symmetries. Since LHC, the future proton-proton collider, is expected to be sensitive to 
scales $ \La $ of the order of a few TeV, the results will be presented
for scales between $ 0.5 $ and $ 50 \tev$.

The paper is organized as follows. In Section~\ref{non_stand}, we define the Lagrangian
relevant for our discussion. Section~\ref{triv_bound} is devoted to the 
derivation of the triviality bound
including effects of non-standard interactions.
In Section~\ref{vac_bound}, we calculate the effective potential with one insertion 
of an effective operator and discuss its
consequences for the vacuum stability bounds.
Concluding remarks are given in Section~\ref{summary}.

\section{Non-Standard Interactions}
\label{non_stand}

Our study of the stability and triviality constraints on the Higgs-boson mass
will be based on the SM Lagrangian modified by the addition of a series
of effective operators whose coefficients parameterize the low-energy 
effects of the heavy physics~\cite{leff.refs}. 
Assuming that these non-standard effects
decouple implies~\cite{decoupling} that the operators appear multiplied
 by appropriate inverse powers of $ \La $.
The leading effects are then
generated by operators of mass-dimension 6 (dimension 5 operators
necessarily violate lepton number~\cite{effe_oper} and are associated with new physics at
very large scales; so we can safely ignore their effects).
Given our emphasis on Higgs-boson physics the effects of all fermions
excepting the top-quark can be ignored~\footnote{We assume that the
masses are natural in the technical sense~\cite{thooft} so that effective
couplings containing the Higgs boson and the light fermions
are suppressed by powers of
the corresponding Yukawa couplings.}. We then have
\bea
\lcal_{\rm tree}& = & 
-\frac14 F_{\mu\nu}^iF^{i\mu\nu} -\frac14 B_{\mu\nu}B^{\mu\nu}+ 
\left| D \phi \right|^2  - 
\lambda \left(-\half v^2 +  |\phi|^2 \right)^2 + \non \\ 
&& i \bar q \not\!\!D q +
i \bar t \not\!\!D t + 
f \left( \bar q \tilde\phi t + \hbox{h.c.} \right) + 
\sum_i\frac{\alpha_i}{\La^2}{\cal O}_i,
\label{lagrangian}
\eea
where $\phi$ ($\tilde \phi = -i \tau_2 \phi^* $), 
$q$ and $t$ are the scalar doublet, third generation 
left-handed quark doublet and the right-handed top singlet, respectively. 
$D$, $F_{\mu\nu}^i$ and $B_{\mu\nu}$ denote a covariant derivative and $SU(2)$, $U(1)$ field strength
whose couplings we denote by $g$ and $g'$.

The factors $\alpha_i$
are unknown coefficients that parameterize the low-energy effects of
the non-standard interactions and we have neglected contributions $
\propto 1/\La^4 $. In addition, for weakly coupled theories, the 
$ \alpha_i $ that can be generated only through loop effects are
sub-dominant as they are suppressed by numerical factors $ \sim 1/
(4\pi)^2 $~\cite{tree_oper}; hence we will consider only those
operators that can be generated at tree-level by the heavy physics. Even
with all the above restrictions there remain 16 operators that 
involve exclusively the fields in (\ref{lagrangian}). Of these only
5 contribute directly to the effective potential, the remaining 11
affect the results only through their RG mixing which, being suppressed by
a factor $ \sim (v/\La)^2 $ are expected to play a sub-dominant role.
In the calculations below we will include only one of these operators;
our results do justify the claim that the corresponding effects are
small.

The following set of operators will be considered:
\beq
\baa{lll}
{\ocal_{\phi}} = \inv3 | \phi|^6 &
{\ocal_{\partial\phi}} = \half \left( \partial | \phi |^2 \right)^2 &
{\ocal_{\phi}\up1} = | \phi |^2 \left| D \phi \right|^2 \cr
{\ocal_{\phi}\up3} = \left| \phi^\dagger D \phi \right|^2 &
{\ocal_{t\phi}} = | \phi |^2 \left( \bar q \tilde\phi t + \hbox{h.c.} \right) &
{\ocal_{qt}\up1} = \half \left|\bar q t \right|^2
\eaa
\label{operators}
\eeq
where
${\ocal_{\phi}}$, ${\ocal_{\partial\phi}}$, ${\ocal_{\phi}\up1}$, ${\ocal_{\phi}\up3}$,
${\ocal_{t\phi}}$ are the 5 operators contributing to the effective
potential, while ${\ocal_{qt}\up1}$ is included to estimate the effects
of RG mixing.

Of the first five operators only  ${\ocal_{\phi}} = \inv3 | \phi|^6$ contributes
at the tree level to the scalar potential:
\beq
V\up{\rm tree}= 
- \eta \Lambda^2 |\phi|^2  + \lambda |\phi|^4  - 
{\alpha_{\phi} \over 3 \Lambda^2 } | \phi|^6
\label{tree_pot}
\eeq
where we have used the notation: $\eta \equiv \lambda v^2/\Lambda^2 $.

\section{Triviality Bound}
\label{triv_bound}

In order to test the high energy behavior of the scalar potential one has to derive
the RG running equations for $\la$, $\eta$ and $\alpha_{\phi}$.
The $\beta$ functions for these parameters are
influenced by all the operators in (\ref{operators})  
and by the gauge and Yukawa interactions, so the full RG evolution also
require the $ \beta $ function for the corresponding couplings.
Both for the $\beta$ functions and then for the effective potential we will
adopt dimensional regularization and $\overline{\rm MS}$ 
renormalization scheme. We will restrict ourself to the
one-loop approximation keeping SM contributions and terms linear in 
the effective operators, defined by 
eq.(\ref{operators}). The evolution equations for the running coupling constants  
are the following: 
\def\dert#1{{ d #1 \over d t}}
\begin{eqnarray}
8 \pi^2 \dert \lambda &=&12\lambda^2 -3 f^4 + 6 \lambda f^2 - 2 \eta
        \left[2 \alpha_\phi + \lambda ( 7 \alpha_{\partial\phi}  +8 \alpha_\phi\up1
        +5 \alpha_\phi\up3)  \right]  \cr
&&+{3\over 2} \left[-\lambda \left(3 g^2 + g'{}^2 \right) 
+ {g'{}^4 +2 g^2 g'{}^2 + 3 g^4\over8}
\right] \cr
8 \pi^2 \dert \eta &=&  \eta\left[6 \lambda + 3 f^2 -2 \eta \left( \alpha_{\partial\phi} 
                + 2\alpha_\phi\up1 +  \alpha_\phi\up3   \right) 
-{3\over 4} \left( 3g^2+ g'{}^2 \right)\right] \cr
8 \pi^2 \dert f &=& {9\over4} f^3 + {1\over2} \eta \left[ 6 \alpha_{t\phi} - f \left(
               \alpha_{\partial\phi} + 2 \alpha_\phi\up1 + \alpha_\phi\up3 + 3
               \alpha_{qt}\up1 \right) \right]  
\cr &&
-{f\over 8} \left( 9 g^2 + {17\over3}g'{}^2 \right) \cr
8 \pi^2 \dert{\alpha_\phi}&=&  
        9 \alpha_\phi \left(6 \lambda + f^2  \right)
  + 12 \lambda^2 (9\alpha_{\partial\phi}+6 \alpha_\phi\up 1
        +5\alpha_\phi\up 3) + 36 \alpha_{t\phi} f^3 \cr
&&- {9 \over 4} ( 3 g^2 + g'{}^2) \alpha_\phi\cr
&& -{ 9 \over 8}
\left[ \alpha_\phi\up1 (3g^4 + 2 g^2 g'{}^2+ g'{}^4) + 
\alpha_\phi\up3 (g^2 + g'{}^2 )^2 \right] \cr
8 \pi^2 \dert{\alpha_{\partial\phi}}&=& 2 \lambda
        \left( 7 \alpha_{\partial\phi} - \alpha_\phi\up1 + \alpha_\phi\up3
              + 3 { \alpha_{\partial\phi} f^2 \over \lambda }
              - 3 { \alpha_{t\phi} f \over \lambda } \right) \cr
8 \pi^2 \dert{\alpha_\phi\up1}&=& 2 \lambda 
        \left(\alpha_{\partial\phi}+5\alpha_\phi\up1 +\alpha_\phi\up3
              + 3 { \alpha_\phi\up1  f^2 \over \lambda }
              - 3 { \alpha_{t\phi} f \over \lambda } \right)  \cr
8 \pi^2 \dert{\alpha_\phi\up3}&=& 6 (\lambda +f^2) \alpha_\phi\up3 \cr
8 \pi^2 \dert{\alpha_{t\phi}} &=& -3 f (f^2+\lambda) \alpha_{qt}\up1 
+ {3\over4} (5f^2-16\lambda) \alpha_{t\phi} \cr
&&- \inv2 f^3 \left( 
        2 \alpha_{\partial\phi} + \alpha_\phi\up1 + \alpha_\phi\up3 \right) \cr
8 \pi^2 \dert{\alpha_{qt}\up1} &=& {3\over2} \alpha_{qt}\up1 f^2
\label{beta_fun}
\end{eqnarray}
where $t\equiv \log (\kappa/\mz)$
and $ \kappa $ denotes the renormalization scale.

From this set of equations it is straightforward to obtain the triviality constraints on
$\mh$ as a function of $\Lambda $ requiring that the position of the Landau pole
is beyond the scale $\La$. 
There is a comment here in order, namely, in actual calculations
the position of the Landau pole cannot be accurately determined to
any finite order in perturbation theory. Therefore the triviality
bound on $\mh$ will be obtained by requiring $\la$ and $\alpha_\phi$ to become
smaller than specified values (as opposed from requiring an actual divergence)
up to the scale $\La$:
\beq
\la(t)\le \la_{\rm max}\;\; {\rm and}\;\; |\alpha_i(t)|\le 1.5\;\; {\rm for}\;\; 0 \le t<\log(\La/\mz). 
\label{triv_con}
\eeq
where we considered $ \lambda_{\rm max} = \pi $ and $ \pi/2 $. 
We have verified that
our results are quite insensitive to the values chosen as upper 
limits for the $ \alpha_i $.

In order to solve the equations (\ref{beta_fun}) we have to specify
appropriate boundary conditions.
For the SM parameters these are determined by requiring that the correct
physical parameters (such as the Higgs-boson and top-quark masses)
are obtained at the electroweak scale.
These initial conditions
should also insure that the correct SM ground state is realized, 
in which the scalar field has the
expectation value $ \langle \pb \rangle \equiv \vtrue  / \sqrt{2} =  246/\sqrt{2} \gev$.
Although  we will discuss the effective potential in more detail later, 
it will be useful to provide here the general 1-loop 
relation between the SM tree-level vacuum
$\vtree$ and the physical electroweak vacuum in the theory defined by the equation 
(\ref{lagrangian}) $\vtrue $ :
\beq
                \vtrue =v + \delta v \;\;\; {\rm for} \;\;\; \delta v \equiv  
                - {1 \over 4 \la(0) v ^2}
                \left.{\partial (V_{\rm eff}-V_{\rm SM}^{\rm (tree)}) 
                \over \partial( \pb/ \sqrt{2} )}\right|_{\pb={v  / \sqrt{2}}},
                \label{vev}
\eeq
where $V_{\rm SM}^{\rm (tree)}$ is the tree-level SM potential and $V_{\rm eff}$
is the 1-loop effective potential that includes effective operator contributions; 
$ \lambda (0) $ denotes the
running coupling constant evaluated at the scale $\kappa=\mz$.
Having the vacuum determined by the above equation, 
the following low-scale relations will
be adopted to fix initial conditions at $\kappa=\mz$ 
for the RG equations for $\la$, $\eta$ and $f$. 
\bea
\mh^2 &=& 2 \la \vtrue ^2\left[ 1 -
{\vtrue ^2 \over 4 \Lambda^2 } \left(4\alpha_{\partial\phi} + \alpha_\phi\up1+ \alpha_\phi\up3+
{2 \alpha_\phi\over\la} \right) \right]+ \mh^{(1)},\cr
\mt &=& {\vtrue  \over \sqrt{2}}\left(f+\alpha_{t\phi}{\vtrue ^2\over \La^2}\right) + \mt^{(1)}
\label{higgs_mass}
\eea
where $ \mh^{(1)},~\mt^{(1)}$ denote the 1-loop radiative corrections to the corresponding masses.
In the calculations bellow we use the expression for $ \mh^{(1)}$ of Ref. \cite{quiros}. 
For the top-quark the deviations from the tree-level value are smaller than the experimental error
and so, for simplicity we will use the expression $\mt= \vtrue f/ \sqrt{2}$. The initial conditions
are non-linear functions of the Higgs-boson mass, and so the solutions to (\ref{beta_fun}) will depend on
both $\Lambda $ and $\mh $.

The boundary conditions for $\alpha_i$ are naturally specified at the scale $\kappa=\La$
since below this scale it is appropriate to describe the effects of the heavy excitations
in terms of the coefficients $\alpha_i$. According to Ref.~\cite{tree_oper} it is natural
to assume that $\alpha_i|_{\kappa=\La} \simeq {\cal O}(1)$.

\begin{figure}[h]                                                    
  \centerline{\epsfxsize=4.5in \epsfbox{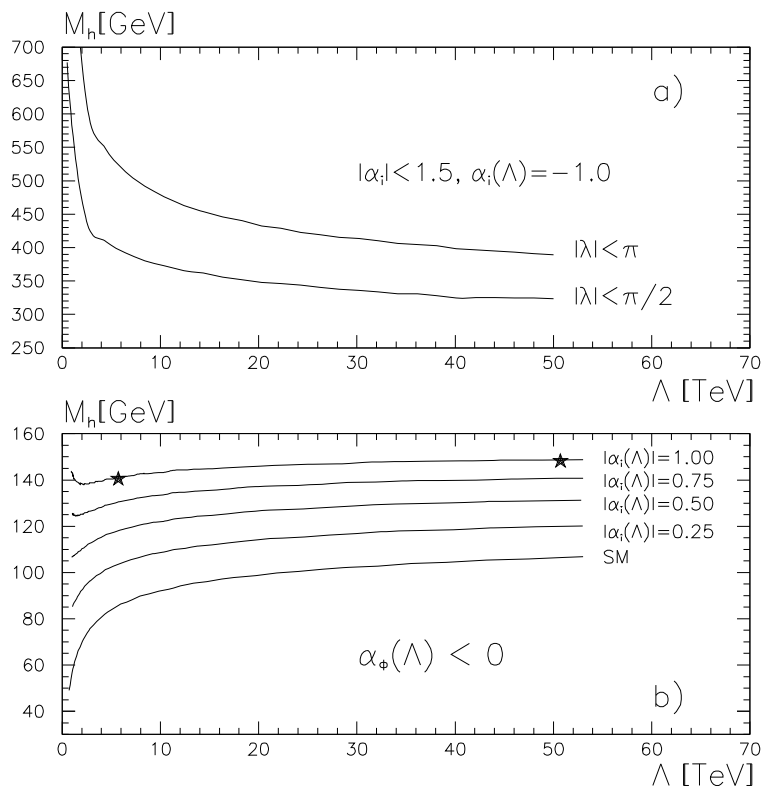}}
  \caption{The upper (a)
(originating from the perturbativity requirement, eq.(\ref{triv_con})),
and the lower (b) (from the condition of the electroweak vacuum stability, eq.(\ref{stab_con}))
bounds on the Higgs-boson mass $\mh$ as a function of the new-physics scale $\La$. 
The lower bounds were obtained for $ \alpha_\phi(\Lambda) < 0 $ and $ \mt=175\gev $.}
\label{bounds}                                                              
\end{figure}

The triviality bound is obtained  by solving the equations (\ref{beta_fun}) with the mixed 
(defined in part at the electroweak scale $\mz$ and at
the new-physics scale $\La$) 
boundary conditions  described above and requiring that at least one of the inequalities 
in eq.(\ref{triv_con}) is
saturated. This provides a relationship between $ \mh $ and $ \La $ that we plot in 
fig.\ref{bounds} (a) for two values of $ \lambda_{\rm max} $.
In order to understand qualitatively the corrections
to the triviality bound we have obtained, is useful to switch off all 
$\alpha_i$ but $\alpha_\phi$. Then,
as it is seen from eq.(\ref{beta_fun}) a Landau pole in the evolution of 
$\la(t)$ causes a singularity
in evolution of $\alpha_\phi$ at {\it this same} energy scale. However, 
as we have just mentioned
it is natural to assume $\alpha_\phi|_{\kappa=\La}\simeq 1$, 
it is clear that strictly speaking it is impossible
to satisfy that condition.
Nevertheless, since we are using a perturbation expansion, we must stop the 
evolution at a scale that
corresponds to a large but finite value $\la_{\rm max}$, 
therefore we can satisfy $\alpha_\phi|_{\kappa=\La}\simeq 1$. 
However, since $d \log \alpha_\phi/ dt $ is positive\footnote{
Here we consider heavy Higgs bosons, therefore $\la$ remains positive in the whole integration region,
it addition $f\gesim g,g'$ what guarantees that $d \log \alpha_\phi/ dt > 0$.}
therefore in the evolution from the scale $\La$ down, $\alpha_\phi(t)$ decreases reaching typically
$10^{-1}\div 10^{-2}$ at the scale $\kappa=\mz$. That explains screening of the effects generated
by operator $\ocal_\phi$: even if $\alpha_\phi|_{\kappa=\La}\simeq 1$ it can not 
grow any larger\footnote{For strongly 
coupled new-physics corrections to this bound see \cite{chan}.}.
So concluding,
the corrections to the SM triviality bound from the non-standard physics 
(embedded in the coefficients $\alpha_i$) are negligible.

\section{Vacuum Stability Bound}
\label{vac_bound}

In order to investigate the vacuum structure of the effective theory we will
first calculate the effective potential:
\begin{equation}
V_{\rm eff} = - \sum_N \inv{N!} \Gamma\up N(0) \pb^n,
\label{eff_pot_def}
\end{equation}
where $\Gamma\up N(0)$ are N-point one-particle-irreducible Green`s functions
with zero external momenta and $\pb$ is the classical scalar field.
Adopting the Landau gauge\footnote{As it has been noticed in Ref.~\cite{gauge_dep_eff_pot}
the effective potential (as a sum of off-shell Greens functions) is gauge dependent.
Therefore the bounds on the Higgs-boson mass 
derived from vacuum stability arguments
can depend on the gauge parameter adopted in the loop calculation~\cite{gauge_dep_bound}. 
However, since the $\beta$
functions and the tree-level potential $V_{\rm eff}\up{\rm tree}$ are 
gauge-independent,
a consistent RG improved tree-level effective potential is in fact gauge independent.
For the one-loop SM RG improved effective potential, the error caused by the gauge 
dependence
has been estimated in Ref.\cite{quiros} at $\Delta \mh \lesim 0.5 \gev$.} 
we obtained:
\begin{eqnarray}
V_{\rm eff}(\pb) &=&
-\eta \Lambda^2 |\pb|^2 + \lambda |\pb|^4 - {\alpha_\phi |\pb|^6\over3 \Lambda^2} \\
&&+ {1 \over 64 \pi^2} \Biggl[ 
   H^2 \left( \ln {H \over \kappa^2} - {3\over2} \right) +
3  G^2 \left( \ln {G \over \kappa^2} - {3\over2} \right) +
6  W^2 \left( \ln {W \over \kappa^2} - {5\over6} \right) \cr
&&+3  Z^2 \left( \ln {Z \over \kappa^2} - {5\over6} \right) -
12 T^2 \left( \ln {T \over \kappa^2} - {3\over2} \right)
- 4\eta^2\Lambda^4 \left( \ln {\eta \Lambda^2 \over\kappa^2} - {3\over2} \right)\Biggr], \non
\label{effpot}
\end{eqnarray}
where 
\begin{eqnarray}
H &=& \lambda (- v^2 + 6 |\pb|^2 ) - \left[\lambda (- v^2 + 6 |\pb|^2 )
(2 \alpha_{\partial\phi} + \alpha_\phi\up1 + \alpha_\phi\up3) + 
5 \alpha_\phi |\pb|^2 \right] {|\pb|^2\over \Lambda^2} \cr
G &=& \lambda (- v^2 + 2 |\pb|^2 ) -  \left[\lambda (- v^2 + 2 |\pb|^2 )
\inv3 (3 \alpha_\phi\up1 + \alpha_\phi\up3)  + \alpha_\phi |\pb|^2
\right] {|\pb|^2\over \Lambda^2} \cr
W &=& {g^2  \over2}|\pb|^2 \left( 1 + { |\pb|^2 \alpha_\phi\up1 \over \Lambda^2 } \right) \cr
Z &=& { g^2 +g'{}^2 \over2} |\pb|^2  \left( 1 + { |\pb|^2 ( \alpha_\phi\up1 + 
\alpha_\phi\up3)   \over \Lambda^2 } \right) \cr
T &=& f^2  |\pb|^2 \left(1 + { 2 \alpha_{t \phi} | \pb|^2 \over f \Lambda^2} \right),\non
\end{eqnarray}
where $g$ and $g'$ denote the $SU(2)$ and $U(1)$ running gauge coupling constants, respectively.
The {\em form} of the effective potential is precisely the same as the
one in the pure SM, the whole effect of the effective operators can be
absorbed in a re-definition of the SM quantities $H$, $G$, etc.\footnote{The same result
(in the leading order in $\alpha_i$)
for the effective potential have been obtained adopting the diagrammatic approach (with
one insertion of an effective operator) according
to eq.(\ref{eff_pot_def}) and also using the functional definition of the effective potential
proposed by Jackiw~\cite{func_def}.}  
It should be noticed here that the last term in eq.(\ref{effpot}) is a constant that
is needed to construct a scale invariant effective potential, 
for details see Ref.~\cite{cosm_const}. The constant term chosen here
is consistent with the diagrammatic definition of the effective potential 
eq.(\ref{eff_pot_def}),
that implies $V_{\rm eff}(\pb=0)=0$.

Since we will consider values of $\pb$ substantially larger then the electroweak
scale $\vtrue $, we shall chose an appropriate renormalization scale 
$\kappa \sim \pb$ in order
to moderate the logarithms that appear in the effective potential.
As in the previous section we shall use the RG running equations
to relate the coupling constants
renormalized at the high scale $\pb$
to the low-scale parameters $\vtrue $, $ \mt$  and $\mh$. 

Finally, (and unlike the pure $\phi^4$) the interaction of the scalars
with the fermions and gauge bosons, generate a non-trivial
scalar field anomalous dimension $\gamma$. We therefore also include the
corresponding scale dependence of $ \pb $:
\beq
\pb(t)=\exp\left\{-\int_{t_z}^t 
\gamma[\la(t'),\eta(t'),f(t'),\alpha_i(t')]dt'\right\} \pb(t_z),
\label{phi_scaling}
\eeq
where 
\beq
\gamma = {1\over 16 \pi^2 }\left[3 f^2 - \frac{9}{4}g^2 - \frac{3}{4}g^{\prime 2}
-   \eta \left(\alpha_{\partial\phi} + 2\alpha_\phi\up1 + \alpha_\phi\up3 \right) 
\right]
\eeq
Hereafter we will consider the RG improved effective potential $ V_{\rm
eff} ( \pb(t)) $.

We note that  the RG improved effective potential given by eq.(\ref{effpot})
is scale invariant. That is, to one loop and ignoring terms quadratic in the $
\alpha_i$, $ V_{\rm eff}$ obeys the renormalization group equation:
\begin{equation}
\kappa\partial_\kappa V_{\rm eff}\up{\rm1-loop} + \left( 
\sum_i \beta_i \partial_{\lambda_i }  - \gamma
\pb \partial_{\pb} \right) V_{\rm eff}\up{\rm tree} = 0 
\end{equation}
where
$V_{\rm eff}\up{\rm tree}$ and $V_{\rm eff}\up{\rm1-loop}$ 
denote, respectively, the tree, eq.(\ref{tree_pot}), and 1-loop, eq.(\ref{effpot}),
contributions to $V_{\rm eff}$, and $\beta_i$ are defined 
in (\ref{beta_fun}).
We note that terms quadratic (and higher) in the $ \alpha_i $ are
associated with contributions of order $ 1/\La^4 $ to the effective
Lagrangian and are sub-dominant.

\begin{figure}[h]                                                    
  \centerline{{\epsfxsize=5in \epsfbox{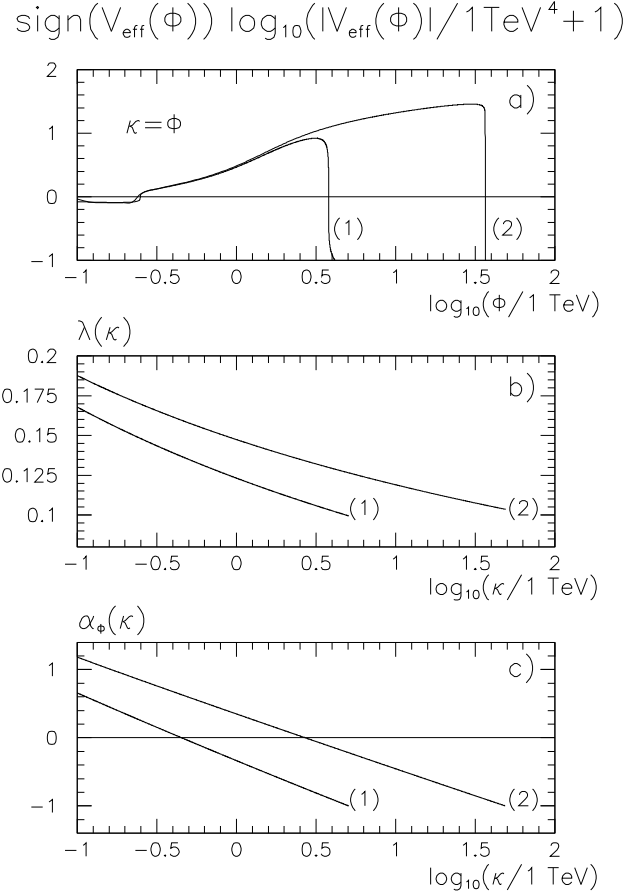}}}
  \caption{The effective potential renormalized at the scale $\kappa=\phi$ (a),
the running of $\lambda$ (b) and $\alpha_\phi$ (c) for the parameter sets (1) and (2) 
defined in the text.}
\label{potential}                                                              
\end{figure}

Fig.\ref{potential} (a) illustrates the behavior of the effective potential
renormalized at the scale $\kappa=\bar{\varphi}$. Since the minimum at 
$\langle \bar{\varphi}\rangle= v_0/\sqrt{2}$
is very shallow, in order to make it visible we plot the following function
of the effective potential:
${\rm sign}(V_{\rm eff})\log_{10}[(V_{\rm eff}/1\tev^4)+1]$.
To show the relevance of RG running of effective-potential parameters
we also plot in Fig.\ref{potential} the evolution of $\la$ (b) and $\alpha_\phi$ (c).
The curves contained in the figure correspond to two sets of initial conditions
(1) and (2) that lead to the Higgs-boson mass and the new-physics scale
marked in Fig.\ref{bounds} by $\star$'s. As it is seen from the figure effects
of the running are substantial, e.g. for the set (2) $\la$ changes by almost 100\%
while $\alpha_\phi$ by more than 200\%. At the electroweak scale,   
$\alpha_\phi$'s start with  positive values, 
however then, through the evolution they switch signs  
and eventually reach $\alpha_\phi=-1$. That should
illustrate the fact that the RG running of the coefficients $\alpha_i$ is crucial
for the stability of the system\footnote{Corrections to the SM vacuum stability bound that emerge
in presence of the operator $\ocal_\phi$ has been previously discussed in Ref.~\cite{datta}.
However, there the authors did not consider one-loop contributions to
the effective potential that are generated by insertions of effective operators.
RG running of $\alpha_\phi$ has also been neglected.}.

The initial conditions for the running couplings
guarantee that the electroweak vacuum is 
at $\langle \pb \rangle=\vtrue /\sqrt{2}$. 
However if $\veff $ at some large value of the field $\pb_{\rm high} $
is smaller than $ \veff(\langle \pb \rangle)$
this vacuum becomes unstable (as there would be a possibility of 
tunneling\footnote{The tunneling time
                                     will not be calculated here, it can be obtained using the procedure 
                                     described in~\cite{tun}; we assume
                                      that it is smaller than the age of the universe.}
towards the region of lower energy).
This will occur when the Higgs-boson mass is sufficiently small
(corresponding to a small value of $ \lambda(0)$), and will provide a lower
bound on $ \mh$. 
In this case $ \pb_{\rm high} $
defines a scale at which the theory breaks down, so that $ \pb_{\rm high} \sim
\Lambda $. In actual calculations we took $ \pb_{\rm high} = 0.75 \Lambda $ 
since (\ref{lagrangian}) is valid for scales below $ \Lambda $, hence
the stability bound on $ \mh$ is determined by the condition
\beq
\veff(\pb =0.75\La)|_{\kappa=0.75\La} =
\veff(\pb = \vtrue /\sqrt{2})|_{\kappa=\vtrue /\sqrt{2}}
\label{stab_con}
\eeq
where, as mentioned previously, we have chosen the renormalization scale
$\kappa$ to tame the effects of the logarithmic contributions to
$\veff(\pb)$. The resulting bound on $ \mh $ as a function of $ \La$
for various choices of $ \alpha_i(\La) $ is plotted in
Fig. \ref{bounds} (b). 

In obtaining the stability bounds of Fig. \ref{bounds} (b) we assumed all
couplings $ \alpha_i $ had the same magnitude at the high scale $
\Lambda $, and $ \alpha_\phi< 0$ (the results are insensitive to the 
sign of the other $ \alpha_i $ except $\alpha_{t\phi}$).
For other values of $\alpha_i$ we
found that when $\La > 300\gev$ there is a curve in the  
$\alpha_{\phi} - \alpha_{t\phi}$
plane below which  either $\pb=174\gev$ is not a minimum or, if it is,
then there is another deeper minimum at a scale
$174 \gev < \pb< 0.75 \La$; we can roughly say
that this unphysical scenario can be avoided if
$\alpha_\phi \lesim -0.1$ ~\footnote{We do not
expect this result to be modified significantly when terms of order
 $ 1/\La^4 $ are included: a contribution $ \sim \alpha\up 8
\pb^8/\Lambda^4 $ can balance the destabilizing effect of $ \ocal_\phi $
only when $ \pb \sim \La $ which again leads to $ \La \sim 300 \gev $.}.

There is an important remark here in order. 
If the Higgs boson mass, as suggested by LEP data,
is indeed~\cite{higgs_disc} $\simeq 115 \gev$, then 
the SM vacuum stability bound implies $ \Lambda \lesim {\cal O}(100) \tev $.
As it is seen in Fig. \ref{bounds} (b) presence of effective operators could dramatically change
the SM picture. Even for the modest values of the coefficient $|\alpha_i|=0.25,~0.50,~0.60$ 
the upper bound on $ \La$ is significantly reduced to   $\La \simeq 20,~4,~1\tev$, 
respectively!

Other limits on the scale $\La$ could be obtained form so called precision observables. 
The most elegant approach is
to calculate the oblique parameters $S$, $T$ and $U$~\cite{stu} within the effective 
theory\footnote{It should be noticed that among operators considered here only $\ocal_\phi^{(3)}$
contributes to the oblique parameters ($T$) 
and therefore is constrained by the precision data, however as it has been 
shown here the operator that is most relevant for the triviality and vacuum stability bound is
$\ocal_\phi$ and contributions from $\ocal_\phi^{(3)}$ are much less important.}
and then fit their experimental values~\cite{stu_effective_1,stu_effective_2}. 
The limits obtained that way depend also on the Higgs-boson
mass $\mh$ therefore it would be interesting to superimpose precision-measurement limits,
the direct LEP limit and 
those obtained here, consistently taking into account higher dimensional operators,
that is however beyond the scope of this paper\footnote{Searches that neglect 
higher-dimensional-operator
corrections to both the triviality and the vacuum stability Higgs-boson bounds 
are published, see Ref.\cite{stu_effective_2}.}. 

\section{Summary and Conclusions}
\label{summary}

We have considered restrictions on the Higgs-boson mass that emerge
form requirement of perturbative behavior of the quartic coupling constant
(the triviality bound) and from the condition of stable electroweak 
vacuum
taking into account possible non-standard interactions described by 
effective operators of dimension $\leq\;6$.   
It was shown that for the scale of new physics in the region 
$\La \simeq 0.5 \div 50 \tev$
the Standard Model triviality upper bound remains unmodified
whereas the lower bound from requirement of vacuum stability is naturally increased
by $40\div60\gev$ depending on the scale $\La$ and strength
of coefficients of effective operators. Therefore the allowed region of 
the Higgs-boson mass is reduced substantially. If the Higgs-boson mass
is close to its lower LEP limit then the upper bound on the scale of new physics that
follows from the vacuum stability requirement could be
decreased dramatically even for modest values of coefficients of effective 
operators implying new physics already at the scale of $\sim 1\div 2\tev$.

\vspace*{0.6cm}
\centerline{ACKNOWLEDGEMENTS}

\vspace*{0.3cm}
This work is supported in part by the State
Committee for Scientific Research (Poland) under grant 5~P03B~121~20
and funds provided by the U.S. Department of Energy under grant No.
DE-FG03-94ER40837. One of the authors (BG) is indebted to
CERN, SLAC and U.C. Riverside for the
warm hospitality extended to him while this 
work was being performed.

\vskip 1cm


\begin{thebibliography}{99}
%
\bibitem{higgs_limit}
T. Junk, The LEP Higgs Working Group, at LEP Fest October 10th 2000,
http://lephiggs.web.cern.ch/LEPHIGGS/talks/index.html.
%
\bibitem{prec_data}
E. Tournefier, The LEP Electroweak Working Group, 
talk presented at the 36th Rencontres De Moriond On Electroweak 
Interactions And Unified Theories, 2001, Les Arcs, France,
hep-ex/0105091. 
%
\bibitem{triviality}
K. Wilson, \xprb{4}{1971}{3184}.
%
\bibitem{triv_bounds}
L. Maiani, G. Parisi, and R. Petronzio, \npb{136}{1979}{115};
M. Lindner, \zfp{31}{1986}{295}.
%
\bibitem{vacuum_bounds}
N. Cabibbo \etal, \npb{158}{1979}{295};
for a review see M. Sher, \pr{179}{1989}{273} and references therein.
%
\bibitem{higgs_disc}
P.~Igo-Kemenes, The LEP Higgs Working Group, at LEPC 3 November 2000,
http://lephiggs.web.cern.ch/LEPHIGGS/talks/index.html.
%
\bibitem{quiros}
J.~A.~Casas {\it et al.} Nucl.\ Phys.\ B {\bf 436}, 3 (1995)
[Erratum-ibid.\ B {\bf 439}, 466 (1995)] [hep-ph/9407389];
M. Quiros, IEM-FT-153-97, hep-ph/9703412.
%
\bibitem{leff.refs} 
S.~Weinberg, hep-th/9702027. 
H.~Georgi, Ann.\ Rev.\ Nucl.\ Part.\ Sci.\ {\bf 43}, 209 (1993). 
%
\bibitem{decoupling}
T.~Appelquist and J.~Carazzone, Phys.\ Rev.\ D {\bf 11}, 2856 (1975).
J.~Collins, F.~Wilczek and A.~Zee, Phys.\ Rev.\ D {\bf 18}, 242 (1978).
%
\bibitem{effe_oper}
W. Buchm\"ueller and D. Wyler, \npb{268}{1986}{621}.
%
\bibitem{thooft}
G.~'t Hooft, 
{\it Lecture given at Cargese Summer Inst., Cargese, France, Aug 26 - Sep 8, 1979},
in {\it C79-08-26.4} PRINT-80-0083 (UTRECHT).
%
\bibitem{tree_oper} 
C. Arzt, M.B. Einhorn and J. Wudka, \npb{433}{1995}{41}, hep-ph/9405214. 
%
\bibitem{chan}
M. Chanowitz, \xprd{63}{2001}{076002}.
%
\bibitem{gauge_dep_eff_pot} 
L. Dolan and R. Jackiw, \xprd{9}{1974}{2904}.
%
\bibitem{gauge_dep_bound}
W. Loinaz and R.S. Willey, \xprd{56}{1997}{7416}, hep-ph/9702321.
%
\bibitem{func_def}
R. Jackiw, \xprd{9}{1974}{1686}.
%
\bibitem{cosm_const}
C. Ford \etal, \npb{395}{1993}{17}, hep-lat/9210033.
%
%
\bibitem{datta}
A. Datta, B.L. Young and X. Zhang, \plb{385}{1996}{225}, hep-ph/9604312.
\bibitem{tun}
%
G.~Isidori {\it et al.}, \npb{609}{387}{2001}, hep-ph/0104016.
%
\bibitem{stu}
M.E. Peskin and T. Takeuchi, \xprl{65}{1990}{964}, \xprd{46}{1992}{381}, see also
G.~Altarelli, R.~Barbieri, \plb{253}{1991}{161}.
%
\bibitem{stu_effective_1}
G. Sanchez-Colon and J. Wudka, \plb{432}{1998}{383}, hep-ph/9805366;
R.~Barbieri and A.~Strumia, \plb{462}{1999}{144}, hep-ph/9905281;
L.~Hall and C.~Kolda, \plb{459}{1999}{213}, hep-ph/9904236;
J.A.~Bagger, A.F.~Falk and M.~Swartz, \xprl{84}{2000}{1385}, hep-ph/9908327.
%
\bibitem{stu_effective_2}
R.S.~Chivukula and N.~Evans, \plb{464}{199}{244}, hep-ph/9907414;
C.~Kolda and  H.~Murayama, \jhep{0007}{2000}{035}, hep-ph/0003170.
%
\end{thebibliography}
\end{document}